# Failure Mode and Effects Analysis (FMEA) for Experimental Use of FLASH on a Clinical Accelerator

Mahbubur Rahman[1*], Rongxiao Zhang[1,2,3*], David J. Gladstone[1,2,3], Benjamin B. Williams[1,2,3], Erli Chen[4], Chad A. Dexter[3], Lawrence Thompson[3], Petr Bruza[1], Brian W. Pogue[1,3,5]

[1] Thayer School of Engineering, Dartmouth College, Hanover NH 03755, US

[2] Department of Medicine, Radiation Oncology, Geisel School of Medicine, Dartmouth College Hanover NH 03755 USA

[3] Norris Cotton Cancer Center, Dartmouth-Hitchcock Medical Center, Lebanon, NH 03756 USA

[4] Cheshire Medical Center, Keene NH 03431 USA

[5] Department of Surgery, Geisel School of Medicine, Dartmouth College, Hanover NH 03755 USA

[*]Correspondence to: Mahbubur.Rahman.TH@dartmouth.edu or Rongxiao.Zhang@hitchcock.org

**Abstract:**

**Background:** Use of a linear accelerator in ultra-high dose rate (UHDR) mode can provide a conduit for wider access to UHDR FLASH effects, sparing normal tissue, but care needs to be taken in the use of such systems to ensure errors are minimized.

**Purpose:** Failure Modes and Effects Analysis (FMEA) was carried out in a team that has been involved in converting a LINAC between clinical use and UHDR experimental mode for more than one year, following the proposed methods of TG100.

**Methods:** A team of 9 professionals with extensive experience were polled to outline the process map and workflow for analysis, and developed fault trees for potential errors, as well as failure modes that would results. The team scored the categories of severity magnitude (S), occurrence likelihood (O), and detectability potential (D) in a scale of 1 to 10, so that a risk priority number (RPN=S*O*D) could be assessed for each.

**Results:** A total of 46 potential failure modes were identified, including 5 with RPN>100. These failure modes involved 1) patient set up, 2) gating mechanisms in delivery, and 3) detector in the beam stop mechanism. Identified methods to mitigate errors included 1) use of a checklist post conversion, 2) use of robust radiation detectors, 3) automation of QA and beam consistency checks, and 4) implementation of surface guidance during beam delivery.

**Conclusions:** The FMEA process was considered critically important in this setting of a new use of a LINAC, and the expert team developed a higher level of confidence in the ability to safely move UHDR LINAC use towards expanded research access.

1. **Introduction:**

FLASH radiotherapy, or radiation treatment with ultra-high dose rate beams (UHDR, >40Gy/s average and >$10^5$ Gy/s intrapulse dose rates) has shown results of reduced normal tissue toxicity[1–6], and so it has seen a resurgence in recent years with many biological experiments and human translation ongoing. One of the major roadblocks recognized in the field is a lack of readily accessible systems that can deliver UHDR beams consistently, and so several institutions have tested existing system modifications for FLASH experiments[7,8]. Proton FLASH beams have been developed recently by modulating the particle flux of their highest energy beams to establish small animal and pre-clinical radiation research platforms[4,9,10]. Modified clinical LINACs have been used to deliver UHDR electron beams, with a reduced source to surface distance for preclinical studies with small animals[7,8]. Dedicated UHDR LINACs exist for experimental use (e.g. Oriatron eRT6, Mobetron)[11–13], and are being applied to for human use, although the approvals and protocols are still evolving today.

In order to investigate the FLASH effect on large animals and future human patients, Rahman et al.[14] modified a clinical LINAC (Varian Clinac 2100 C/D) to deliver UHDR beams at the isocenter with minimal changes to the clinical geometry. The method followed the previous work of Schüler et al[7] and Lempart et al[8], and required retraction of the photon target, and allowing the electron beam with the machine running in photon mode,) to pass through an empty carousel port, permitting the use of all the clinical accessories such as collimating cones. Since its first conversion the machine has been utilized to treat mice, minipigs, and several spontaneous canine tumor patients[15]. Nonetheless Newell et al. raised concerns about the clinical machine's use for FLASH experiments and how the conversion or modifications could potentially cause mistreatments[16]. As part of the concern around use of such as novel system a detailed analysis of safety was undertaken as described here.

The medical physics community has recently developed methods to identify and mitigate such failures and established risk based quality management programs via failure modes and effects analysis (FMEA)[17]. The FMEA process is a method to identify possible failures in the design and use of a system and combine this with the effects of these failures on the use outcome[18]. Since its adoption by the US military[19] in the 1940's it has been utilized in many industries, including extensively in healthcare[20]. The medical physics community developed recommendations in AAPM task group report 100 to implement this risk management technique[17] and FMEA has been performed by many institutions to mitigate errors of treatment modality programs[21–25] as well as new or existing technology[26–28].

In this study, to mitigate potential errors from conversion and use of a clinical LINAC for FLASH experiments, an FMEA study was conducted with a team of 9 professionals following the methodologies described in AAPM TG 100 report. This FMEA study was conducted during performing FLASH experiments over the course of more than one year, with safe conversions between FLASH and conventional modes approaching 100 times. Methods for mitigating errors and improving the FMEA process were also discussed.

## 2. Materials and Methods:

### 2. A. Organization of Multidisciplinary Team

A multidisciplinary team consisting of four clinical physicists, two biomedical engineers, two physics/engineering researchers, and one senior PhD student provided input for this FMEA. The team was intentionally diverse to include professionals' experience with the clinical LINAC under FLASH and conventional conditions. The four clinical physicists had first-hand experience utilizing the clinical LINAC for patients and one of the physicists had first-hand experience with delivering the FLASH beam for experiments and converting the LINAC. The two researchers developed technology and experiments that investigate the FLASH dosimetry and radiochemistry effects. The two factory trained biomedical engineers had the most technical experience in converting the LINAC and ensuring consistency in beam delivery. The PhD student had the most experience validating and delivering the beam under FLASH conditions and conducting experiments (dosimetry for in vivo, in vitro, and quality assurance). The team scheduled weekly meetings for approximately 4 weeks to initially develop the process maps, workflow, fault trees, and potential failure modes followed by approximately 3 weeks of meetings to estimate scores and suggest methods of mitigating errors. While there was at least one scheduled meeting each week there were other opportunities throughout other meetings/discussions that led to further suggestions in mitigating error, potential failures etc.

## 2. B. Process Map and Workflow

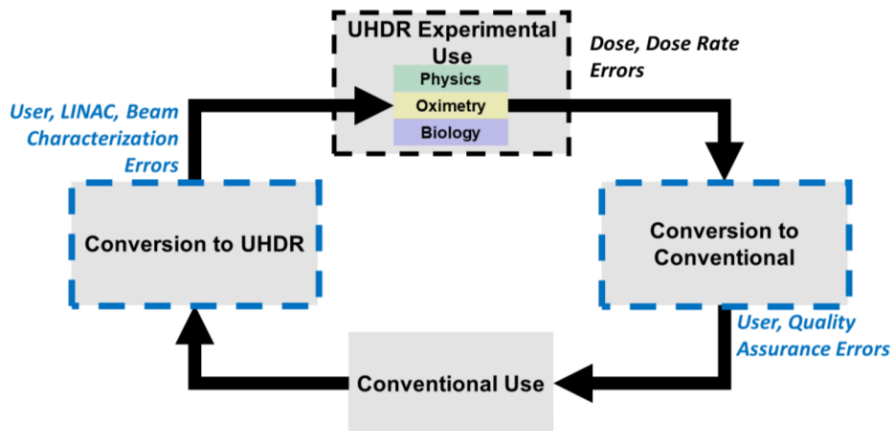

**Figure 1.** Workflow of FLASH/UHDR experiments considering conversions to FLASH mode and then back to conventional mode. Text highlighted in blue are process map steps for conversion and text highlighted in black include fault tree steps with dose and dose rate errors during experimental use.

The FMEA was conducted for the LINAC following the methodology described in AAPM TG 100 report[17]. **Figure 1** demonstrates the workflow of the clinical LINAC's (Varian Clinac 2100 C/D) use under FLASH and conventional condition as well as its conversions. The scope of the analysis was defined for conversion of the machine and during experimental use under FLASH conditions. The process and failures during conventional use were not considered as the manufacturers of the machine have interlocks and features in place to ensure safe and consistent delivery of the beam. The process maps, shown in **Figure 2** and **Figure 5**, were developed for the conversion of the machine to FLASH and back to conventional with the portion of the team responsible for conversion and its use under FLASH conditions. The process maps, adapted from Rahman et al[14], included the modification made to the LINAC as well as add-ons (e.g. dose controller, quality assurance measurements) needed to ensure successful conversion of the machine[29,30]. The team also distinguished between human error and the machine/detector errors.

## 2. C. Fault Trees

Faults trees were created for the use of the machine under FLASH condition with the portion of the team responsible for FLASH experiments. The types of errors considered were dose errors, **Figure 3**, and dose rate errors, **Figure 4**. The potential causes were organized by logic gates and the types of faults were also discussed such as human, quality assurance, or machine error.

## 2. D. Failure Modes

After constructing process maps and fault trees, failure modes for each task in the conversion and cause of the beam delivery faults during FLASH irradiation were defined as shown in **Tables 2-4**. The failure modes' effect upon the results or experimental outcomes and causes of the failure were also established. For the process maps, the errors were categorized by parts of the machine, and QA tool responsible for the error (e.g. pulse counter error). From the dose and dose rate fault trees the fundamental failure modes for the experimental use of the machine were defined.

## 2. E. Scoring and Risk Priorities

The scoring was defined based on an adaptation of the TG100 report scoring guide (**Table 1**), which included scoring of Severity (S), Occurrence (O), and Detectability (D) with a scale of 1 to 10 and risk priority number (RPN=S*O*D). In this study, the scoring differentiated severity for in vivo, in vitro, and quality assurance tool experiments as well as consideration of dose rate differences compared to intended. As shown in **Table 2-4**, the weighted average and the standard deviation in the scores were calculated from the scores determined in the team sessions. During the scoring and afterwards, separate discussions of the team allowed for suggestions for mitigating errors with an RPN >100.

## 3. Results:

The FMEA team considered 46 total potential errors during the conversions and the experimental use of the LINAC. As shown in **Figure 2** and **Table 2**, 24 potential failure modes were defined from the conversion of the beam to UHDR and characterization for use in FLASH experiments. The processes defined were based on experiences of the team with the LINAC from conversion of nearly 100 times over the course of more than a year and prior studies for conversion[14], developing beam controlling[29], and characterization methods[30]. The team generalized the errors to include errors that may arise from any beam controlling and beam characterization methods while incorporating the institution's method of converting the LINAC to deliver FLASH beams. For conversion of the LINAC the potential errors involved were with the carousel positions, target positions, energy switch positioning, servos errors, and energy selection error. In terms of methods of controlling dose, the potential errors were categorized as the beam gating, pulse counter, or dose detection errors. For beam characterization and verifying consistency of the beam, the identified errors included areas such as pulse counting, phantom consistency, dose profile detector and LINAC errors.

The FMEA team identified 9 potential failure modes during the delivery of UHDR beam for FLASH experiments as shown in **table 2**. The processes were regarding predominately the setup of the sample/animal, inputting desired beam parameters, dose, delivery, and its confirmation. Fault tree analysis was completed for considerations of both dose, **Figure 3**, and dose rate, **Figure 4**. The team identified that there could either be under or overdose of the intended or prescribed dose for the experiments. While the fault tree shows many potential errors that could results in under or overdosing, the overall reasons include errors with the LINAC beam parameters, dose controller /beam gating, beam profile consistency check, and set-up of the animals or samples. The fault tree for dose rate was created for the situation when the dose rate is less than 40 Gy/s, as literature currently does not suggest an upper threshold on dose rate for the FLASH effect and it was assumed that the machine was optimized for UHDR delivery. Again, the underlying reasons for the errors involve change in beam parameters, dose controller, beam profile consistency checks, and set up of samples/animals.

The team also identified 13 potential failure modes while converting the LINAC back to deliver conventional beams, as shown in **Figure 5** and **Table 4**. In comparison to converting the machine to deliver UHDR, there were fewer failure modes, as the LINAC software systems are designed to identify potential errors once the servos are turned back on, and so this internal proprietary software checking operational errors encountered in normal clinical mode were considered beyond the scope of this study. Nonetheless, the team assumed that performance validation would always be done after conversion, and so considered errors that might arise while ensuring the conventional beam parameters were consistent with previous measurements under conventional conditions. To convert the LINAC back to conventional mode the team identified that it was a matter of letting the target, energy switch, carousel port go back to its desired position once the air valves were returned to conventional mode.  Furthermore, while under FLASH conditions, special dosimetry considerations were made such as high temporal resolution detectors, under conventional conditions clinical quality assurance tools (e.g. daily QA phantom) can be utilized to verify beam parameters.  Still, errors could arise in returning the machine back to conventional mode which includes carousel, target, energy switch, servos, energy selection, dose profile detector, and LINAC errors.

While scoring each of the failure modes for severity, occurrence, detection, and RPN number, the team suggested methods of mitigating errors for RPN>100.  5 potential failures (highlighted in black in **Table 2, 3** and **4**) with an RPN >100 included:

1. The gating mechanism not turned on while setting up the dose /pulse controller
2. Inconsistent or incorrect calibration of the dose/pulse controller
3. Misalignment of the samples with respect to intended set up
4. Verification of treatment set up including couch alignment and LINAC geometry
5. Second Verification of the sample/patient set up at the treatment console on the camera

Some suggestions to mitigate these errors included implementation of checklists during the conversion processes following some strategies discussed in AAPM task group 275[31]. Furthermore, the checklist should be reimplemented each time there are modifications made to the LINAC or detectors for monitoring the beam delivery. Aside from human error, calibration of the dose/pulse controller not being accurate may be caused by deterioration of the detector from irradiation under FLASH conditions like that described in Ashraf et al[29]. Thus, the team suggested investigating alternative detectors for dose control that demonstrate minimal radiation damage and decrease in sensitivity. Another suggestion was stopping the beam with a threshold for dose and pulses simultaneously as a second measure to ensure the beam does not overdose significantly (>20%) during experimental use (assuming consistency in dose per pulse during delivery). To mitigate errors in misalignment of patient and verifying set up patient or sample the team suggested imaging prior to or during beam delivery ideally at high temporal resolution. Some technology suggested included the Varian On-Board Imager® (OBI) System that image via kV x-rays, Vision RT surface guidance radiation therapy tools, or DoseOptics LLC Cherenkov surface dose imaging technology. It is worth noting there were no failure modes with RPN>100 following conversion back to conventional mode.

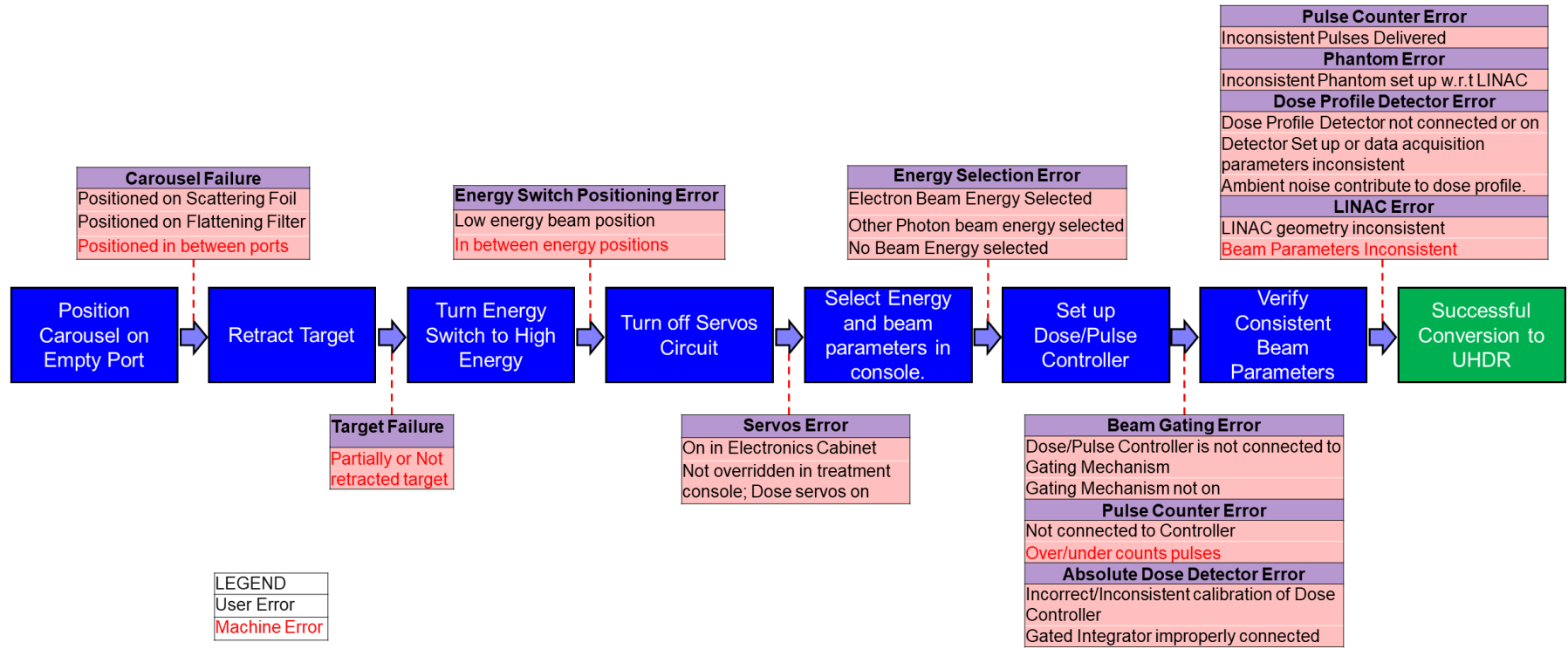

**Figure 2.** Process tree steps (blue boxes) for converting the clinical LINAC to deliver UHDR beam, including the potential errors (pink boxes). The errors in black font are user input errors as they take steps to convert the machine, while errors in red font are machine errors (e.g. LINAC, dose controller, pulse controller).

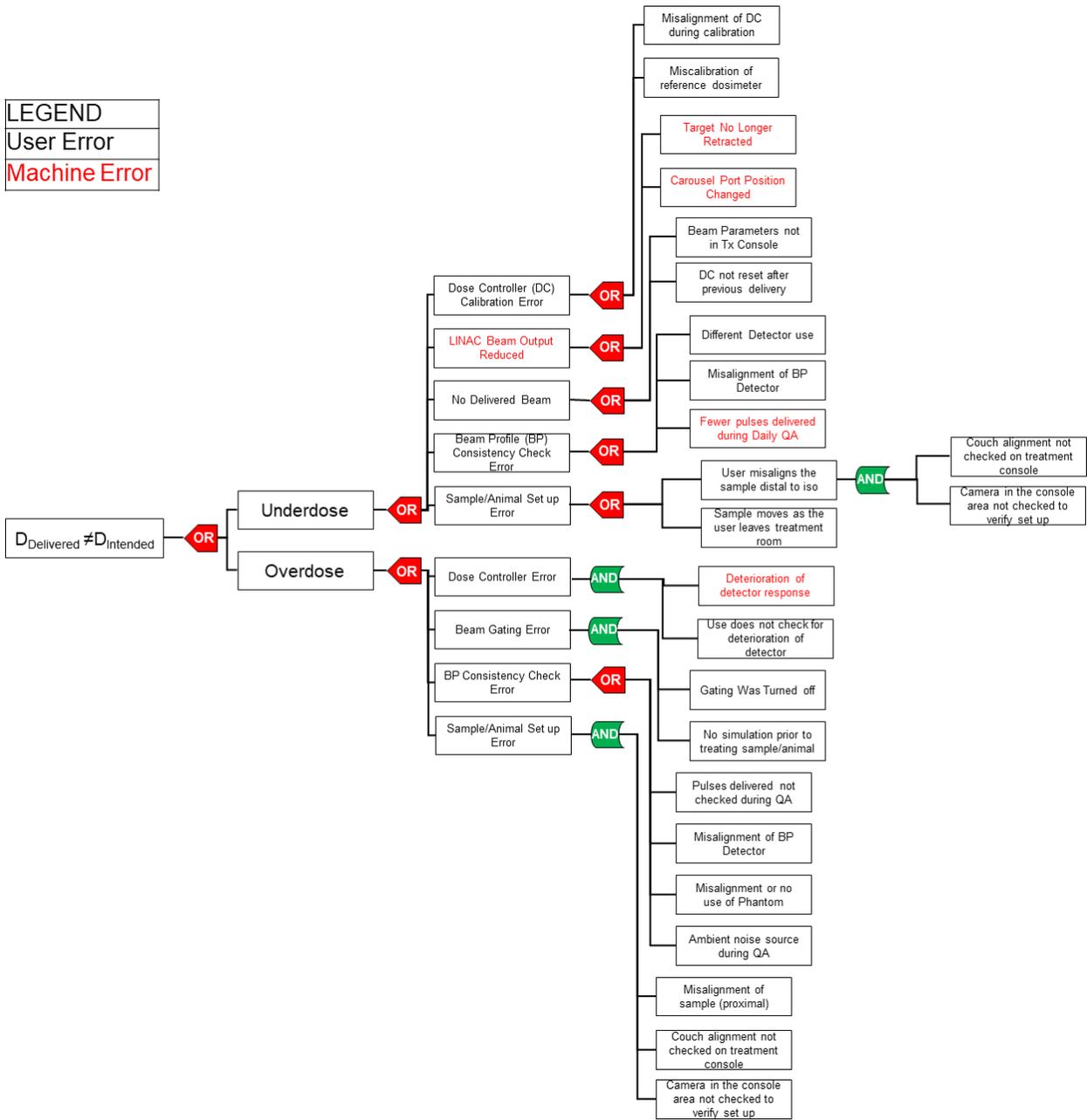

**Figure 3.** Fault tree analysis for UHDR experimental use of a LINAC with considerations of dose. The errors in black font are user input errors, while errors in red font are machine errors (e.g. LINAC, dose controller, pulse controller).

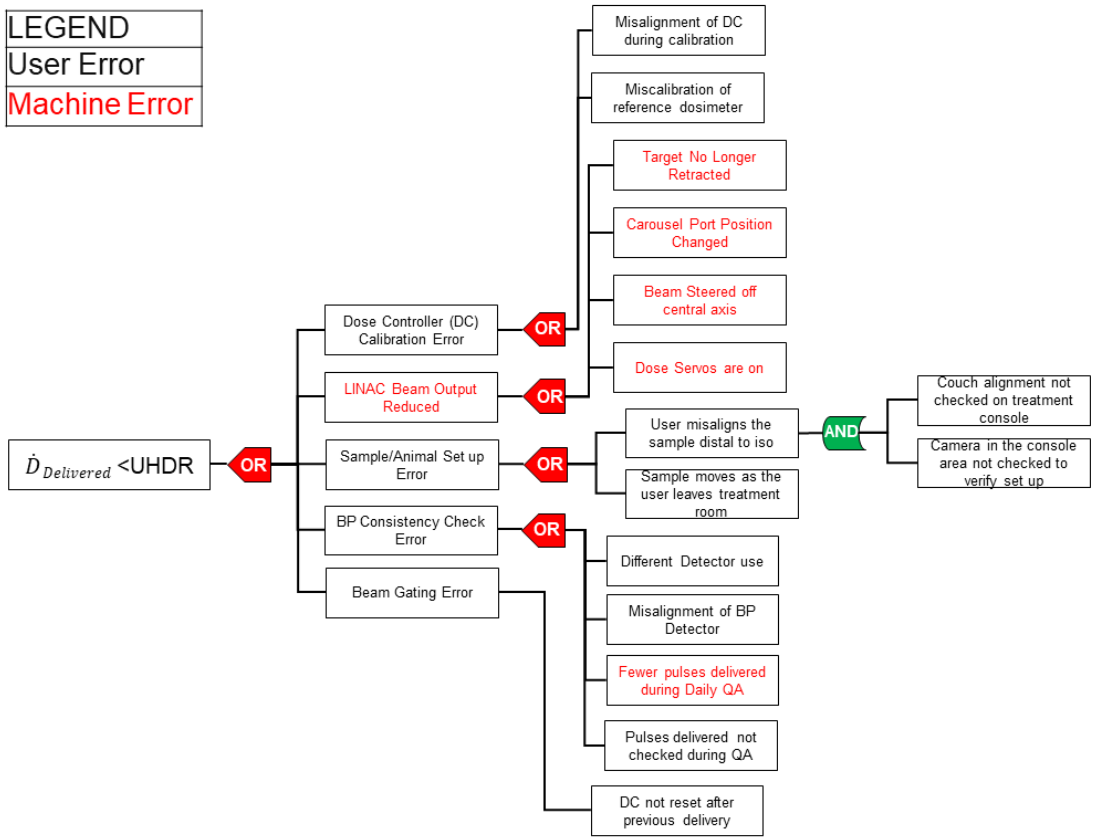

**Figure 4.** Fault tree analysis for UHDR experimental use of a LINAC with considerations of dose rate. The errors in black font are user input errors while errors in red font are machine errors (e.g. LINAC, dose controller, pulse controller).

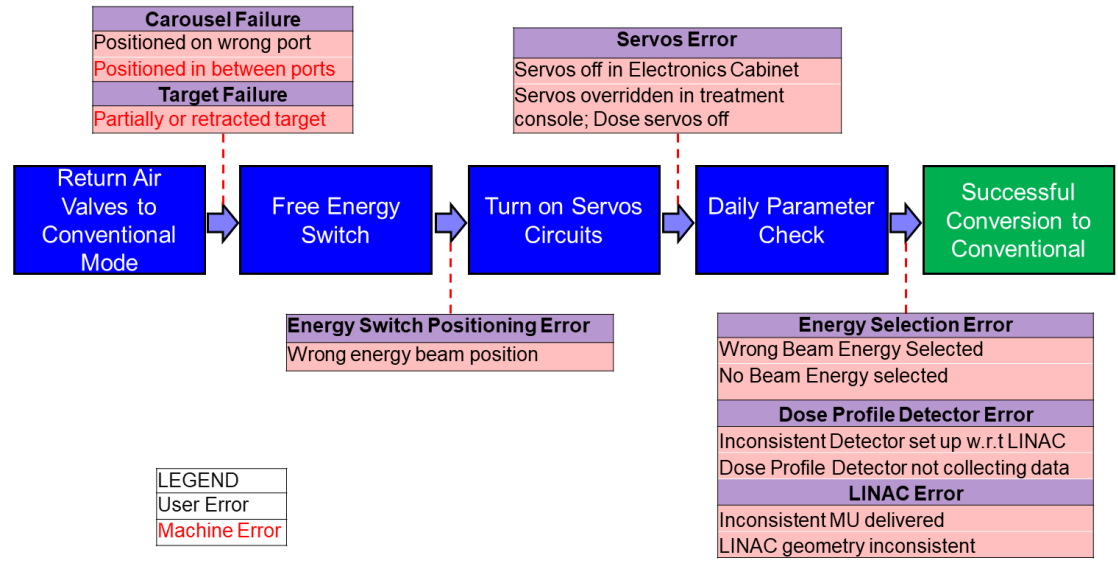

**Figure 5.** Process tree steps (blue boxes) for converting the clinical LINAC back to conventional mode, including the potential errors (pink boxes). The errors colored in black font are user input errors as they take steps to convert the machine while errors in red font are machine errors (e.g. LINAC, dose controller, pulse controller).

| Score | Severity (S) | | Score | Occurrence (O) | | Score | Detectability (D) | |
|---|---|---|---|---|---|---|---|---|
| | Harm to Patient or Experiment | Outcome of Failure | | Description | Frequency (%) | | Description | Probability of Failure going undetected (%) |
| 1 | No effect | Unlikely $D/\dot{D}$/pos. error | 1 | Failure Unlikely | 0.01 | 1 | Always detectable via 1+ methods | 0.01 |
| 2 | No Side Effect | Minimal $D/\dot{D}$/pos. error | 2 | | 0.02 | 2 | Easily detectable via 1+ methods method | 0.2 |
| 3 | Minimal Side Effect | | 3 | | 0.05 | 3 | | 0.5 |
| 4 | Minor Harm – No Side Effects | Minor $D/\dot{D}$/pos. error | 4 | Relatively Few Failures | 0.1 | 4 | Moderately detectable via 1+ methods | 1 |
| 5 | Minor Harm – Minor Side Effects | Minor $D/\dot{D}$/pos. error– In Vitro/QA: $\Delta D, \Delta \dot{D} > 20\%$ | 5 | | <0.2 | 5 | | 2 |
| 6 | | Minor $D/\dot{D}$/pos. error– In Vivo: $\Delta D, \Delta \dot{D} > 5\%$ | 6 | Occasional Failure | <0.5 | 6 | | 5 |
| 7 | Major Harm – serious side effects | Major $D/\dot{D}$/pos. error -In Vitro/QA: $\Delta D, \Delta \dot{D} > 50\%$ | 7 | | <1 | 7 | Difficult to detect via 1+ methods | 10 |
| 8 | | Major $D/\dot{D}$/pos. error -In Vivo: $\Delta D, \Delta \dot{D} > 20\%$ | 8 | Repeated Failure | <2 | 8 | | 15 |
| 9 | Major harm - life threatening | Severe $D/\dot{D}$/pos. error | 9 | | <5 | 9 | Very Difficult to detect via 1+ methods | 20 |
| 10 | Death | Catastrophic $D/\dot{D}$/pos. error | 10 | Failure inevitable | >5 | 10 | Never detectable via 1+ methods | >20 |

**Table 1.** Scoring of severity (S), occurrence (O), and detection (D) to calculate risk priority number (RPN=S*O*D) of the LINAC's use for conversion to UHDR, experimental use in UHDR, and conversion back to conventional. Adapted from TG 100.

| | | Conversion to UHDR | | | | | |
|---|---|---|---|---|---|---|---|
| Process Step/Input | Potential Failure Mode | Potential Failure Effects | Potential Causes | Severity | Occurrence | Detection | Risk Priority Number (RPN) |
| Position Carousel on Empty Port | Carousel Positioned on Scattering Foil | Uniform beam field and reduce dose rate (~5x) at isocenter | Appropriate carousel position not chosen properly | 7.4±0.5 | 2.7±0.1 | 2.0±0.4 | 40.3±9.3 |
| | Carousel Positioned on Flattening Filter | Significantly reduced dose rate, unintended Bremsstrahlung production | Appropriate carousel position not chosen properly | 7.4±0.5 | 2.7±0.1 | 2.0±0.4 | 40.3±9.3 |
| | Carousel Position in between ports | Uncharacterized beam profile; unintended Bremsstrahlung production | Air drive malfunction | 5.4±1.0 | 1.3±0.3 | 1.4±0.4 | 10.0±3.9 |
| Retract Target | Target does not or only partially retract | Conventional x-ray beam will be produced (~0.1Gy/s) | Installed Air Drive not in position to control the target air drive. Failure of Humphrey air valve. | 6.6±0.9 | 2.0±0.4 | 3.4±0.3 | 45.1±12.1 |
| Turn Energy Switch to High Energy | Energy switch in low energy beam position | Electron UHDR beam will have a different energy, and dose rate | Installed air drive was not positioned for high beam energy. Failure of Humphrey valve | 6.3±1.0 | 2.6±0.3 | 1.7±0.4 | 27.7±8.3 |
| | Energy Switch in between energy positions | No beam output? | Installed air drive was not positioned for high beam energy. Failure of Humphrey valve | 6.3±1.0 | 2.0±0.4 | 1.7±0.4 | 21.6±7.6 |
| Turn off Servos Circuits | Servos on in Electronics Cabinet | Beam output reduces and flatness and symmetry adjusted over multiple deliveries | Servos adjusts beam to match expected symmetry and dose rate of conventional beam - Forgot to turn off servos | 4.9±0.8 | 4.4±0.3 | 3.7±0.5 | 79.9±18.8 |

| Step | Failure Mode | Effect | Cause | O | S | D | RPN |
|---|---|---|---|---|---|---|---|
| | Servos not overridden in treatment console; Dose servos on | Beam output reduces and flatness and symmetry adjusted over multiple deliveries | Servos adjusts beam to match expected symmetry and dose rate of conventional beam | 4.9±0.8 | 4.6±0.3 | 3.7±0.5 | 82.5±19.1 |
| Select Energy and beam parameters in console. | Electron (9MeV, 12 MeV, etc.) Beam Energy Selected | Conventional electron beam will be produced but with a gaussian shape at sub-FLASH dose rates | Human Error | 8.0±0.9 | 2.2±0.2 | 1.2±0.2 | 21.1±4.3 |
| | 6MV beam energy selected | Beam Energy and output different from 10MeV UHDR beam? | Human Error | 8.4±0.5 | 2.0±0.0 | 2.8±0.2 | 47.0±4.3 |
| | No Beam Energy selected | No beam produced | User error | 1.6±0.5 | 2.0±0.0 | 1.2±0.2 | 3.8±1.4 |
| Set up Dose/Pulse Controller | Dose/Pulse Controller is not connected to Gating Mechanism | Beam will deliver MU number from LINAC | User did not connect | 7.2±0.7 | 2.4±0.4 | 1.6±0.5 | 27.6±10.5 |
| | Gating Mechanism not on | Beam will deliver MU number from LINAC | User did not turn switch | 7.2±0.7 | 6.4±0.5 | 6.8±1.1 | 313±64 |
| | Pulse Counter not connected to Controller | Dose Controller will not readout pulses delivered and stop beam delivery. | User did not connect power source, or cables | 3.2±0.2 | 2.4±0.4 | 1.2±0.2 | 9.2±2.0 |
| | Pulse Counter over/under counts pulses | measured dose by controller or BP Daily QA will be inconsistent to baseline | Stray radiation effects the pulses counted; counter does not distinguish pulses delivered properly | 4.4±0.4 | 2.4±0.4 | 6.4±0.5 | 67.6±12.8 |
| | Incorrect/Inconsistent calibration of Dose Controller | BP Daily QA will be inconsistent to baseline | Deterioration of the detector, misalignment of the phantom, miscalibration of the reference dosimeter | 6.2±0.2 | 6.4±0.5 | 4.4±0.4 | 175±21 |

| Process Step | Failure Mode | Effect | Cause | O | S | D | RPN |
|---|---|---|---|---|---|---|---|
| | Gated Integrator improperly connected | Dose Controller will not run, and beam will not stop. BP Daily QA will be inconsistent to baseline | Integrator power cable not connected; cable channels switched | 8.6±0.4 | 1.6±0.5 | 1.6±0.5 | 22.0±10.5 |
| Verify Consistent Beam Parameters | Inconsistent Pulses Delivered | BP Daily QA will be inconsistent to baseline | pulse counter under/over counts pulses | 6.0±0.0 | 2.6±0.5 | 4.0±0.0 | 62.4±12.9 |
| | Inconsistent Phantom set up w.r.t LINAC | BP Daily QA will be inconsistent to baseline | Phantom not aligned at 100 cm SSD, orthogonal to beam C.Ax., enough backscatter (5 cm thickness) | 4.4±0.4 | 6.6±0.4 | 2.0±0.9 | 58.1±26.6 |
| | Dose Profile Detector acquisition not occurred | The Daily QA program will not run | Dose Profile Detector not connected or on | 4.4±0.0 | 6.6±0.4 | 1.2±0.2 | 34.8±6.2 |
| | Detector Set up or data acquisition parameters inconsistent | BP Daily QA will be inconsistent to baseline | exposure, gain, triggering parameters in consistent | 4.0±0.0 | 9.0±0.9 | 2.0±0.9 | 72.0±33.0 |
| | Ambient noise contributes to dose profile. | BP Daily QA will be inconsistent to baseline | room light left on (e.g., scintillation-based profile imaging) | 4.0±0.0 | 8.8±1.1 | 2.0±0.9 | 70.4±32.6 |
| | LINAC geometry inconsistent | BP Daily QA will be inconsistent to baseline | e.g., LINAC not at 100 cm SSD, pointed orthogonal to couch surface | 4.0 | 6.4±0.5 | 1.2±0.2 | 30.7±5.3 |
| | Beam Parameters Inconsistent | BP Daily QA will be inconsistent to baseline | Beam steering on, target not retracted, carousel position not in empty port | 4.4±0.4 | 2.4±0.4 | 1.2±0.2 | 12.7±2.9 |

**Table 2.** Failure modes and error analysis of conversion of LINAC to deliver UHDR. (BP=Beam Profile). RPN numbers shaded in black are errors with a score of 100 or greater. All scores include the mean and standard deviation from the surveyed committee members.

| | | UHDR Experimental Use | | | | | |
|---|---|---|---|---|---|---|---|
| Process Step/Input | Potential Failure Mode | Potential Failure Effects | Potential Causes | Severity | Occurrence | Detection | Risk Priority Number (RPN) |
| Set up the sample/animal on the treatment couch | User misaligns the sample w.r.t. isocenter | Wrong dose delivered | sample distal or proximal to intended SSD | 8.2±0.2 | 6.2±0.7 | 4.0±0.0 | 203±24 |
| | Sample moves as the user leaves the room | Wrong dose delivered and not delivered on intended site | Sample position change | 8.0±0.0 | 2.0±0.0 | 4.0±0.0 | 64.0±0.0 |
| | LINAC geometry is not as intended. | Wrong dose delivered | User forgot to change LINAC geometry to intended. | 8.0±0.0 | 2.2±0.2 | 4.2±0.2 | 73.9±6.8 |
| Reset Dose Controller/Beam Stopping Mechanism | User does not reset dose controller or beam stopping mechanism. | Delivers based on MU from LINAC (overdose) | DC dose not gate beam properly | 7.8±0.2 | 3.2±0.2 | 2.8±0.2 | 69.9±6.1 |
| Set delivered MU/beam parameters on treatment console | Wrong MU/beam parameter put in treatment console | underdose of sample/animal | user error in putting in correct parameters | 6.6±0.5 | 2.8±0.2 | 2.8±0.2 | 51.7±6.3 |
| Verify treatment set up and delivery Parameters | Couch alignment/LINAC geometry not checked on treatment console | misaligned sample/animal w.r.t. to prescribed or intended | animal/sample may have moved Human error | 8.0±0.0 | 6.4±0.5 | 8.0±0.9 | 410±57 |
| | Camera in the console area not | misaligned sample/animal w.r.t. to prescribed or intended | animal/sample may have moved Human Error | 8.0±0.0 | 2.2±0.2 | 8.0±0.9 | 141±20 |

| Process Step/Input | Potential Failure Mode | Potential Failure Effects | Potential Causes | Severity | Occurrence | Detection | Risk Priority Number (RPN) |
|---|---|---|---|---|---|---|---|
| | | checked to verify set up | | | | | |
| Deliver treatment | LINAC does not deliver beam | Slows down experiment and user must check to see source of no beam delivery. Troubleshooting. | Beam gating malfunction, DC not reset, beam parameters not inputted into the treatment console of LINAC | 2.4±0.4 | 7.0±0.9 | 1.2±0.2 | 20.2±5.0 |
| Confirm Delivery of Treatment | Secondary dose checks inaccurate | Response of the sample/animal will be not as expected | Dosimeter is miscalibrated, Dosimeter moved from site of interest | 3.8±0.7 | 1.6±0.5 | 2.4±1.3 | 14.6±9.5 |

**Table 3.** Failure modes and error analysis of UHDR Delivery for experimental use (phantoms, in vivo, and in vitro). ). RPN numbers shaded in black are errors with a score of 100 or greater. All scores include the mean and standard deviation from the surveyed committee members.

| Conversion to Conventional ||||||||
|---|---|---|---|---|---|---|---|
| Process Step/Input | Potential Failure Mode | Potential Failure Effects | Potential Causes | Severity | Occurrence | Detection | Risk Priority Number (RPN) |
| Return Air Valves to Conventional Mode | Positioned on wrong port | Interlock prevents delivery | Wrong scattering foil or flattening filter used to create homogeneous field | 7.1±1.2 | 1.0±0.0 | 1.1±0.1 | 8.2±1.7 |
| | Positioned in between ports | Interlock prevents delivery | The beam hits part of the carousel instead of a scattering foil/flattening filter | 7.1±1.1 | 1.0±0.0 | 1.1±0.1 | 8.2±1.7 |
| | Partially or retracted target | Interlock prevents delivery | No bremsstrahlung created for x-ray beams, or electron beam partially hits target | 6.6±0.9 | 1.6±0.2 | 1.1±0.1 | 11.8±2.7 |
| Free Energy Switch | Wrong energy beam position | Wrong beam energy produced by LINAC, | User did not free energy switch or LINAC did not | 5.7±0.9 | 1.1±0.1 | 1.4±0.4 | 9.3±3.2 |

| | | (no resonance in waveguide?) | respond to energy switch in treatment console | | | | |
|---|---|---|---|---|---|---|---|
| Turn on Servos Circuit | Servos switch off in Electronics Cabinet | Interlock prevents delivery | User did not turn on Servos | 6.1±1.0 | 4.7±0.9 | 1.1±0.1 | 33.1±9.0 |
| | Servos overridden in software; Dose servos off | Machine will not correct any change in output, profile. | User did not reset the console to remove all the overridden interlocks, User did not turn on dose servos in console | 8.1±0.1 | 2.4±0.7 | 2.3±0.5 | 45.2±16.6 |
| Verify Daily Parameter Check | Wrong Beam Energy Selected | Energy not matched in QA check | Human Error | 2.0±0.3 | 6.0±0.6 | 1.1±0.1 | 13.7±3.1 |
| | No Beam Energy selected | No beam is delivered | Human Error | 2.0±0.3 | 6.0±0.6 | 1.1±0.1 | 13.7±3.1 |
| | Inconsistent Detector set up w.r.t LINAC | QA output, symmetry, or flatness consistency failure. | Human error | 2.8±0.7 | 6.2±0.7 | 1.2±0.2 | 20.8±6.6 |
| | Dose Profile Detector not collecting data. | No data. | User did not properly connect the detector to a power source, or interface to gather/analyze the data. | 2.4±0.4 | 6.2±0.7 | 1.2±0.2 | 17.9±4.3 |
| | Inconsistent MU delivered | Beam output measure is inconsistent | User inputted wrong number of MUs in treatment console. | 2.4±0.4 | 6.0±0.9 | 1.2±0.2 | 17.3±4.5 |
| | LINAC geometry inconsistent | Beam angled or different distance from the detector | User did not set the LINAC in the same position regarding SSD, collimator rotation, couch position, gantry rotation | 2.4±0.4 | 6.2±0.7 | 1.6±0.5 | 23.8±9.2 |

**Table 4.** Failure modes and error analysis of conversion to conventional beam. RPN numbers shaded in black are errors with a score of 100 or greater. All scores include the mean and standard deviation from the surveyed committee member

## 4. Discussion

After this institutional FMEA, team members are now stronger proponents of future risk analysis of other treatment modalities in the radiation oncology suite because this study was found to provide a logical objective framework to mitigate risk for FLASH irradiators. The experiences of the clinical physicists were crucial to improving the current FLASH experimental methods, particularly since historically they witnessed many errors during treatment of humans with a greater sense of severity, occurrence, and detectability from a clinical perspective. Their suggestions addressed predominantly patient positioning errors and the beam monitoring/controlling errors that may occurring during conversion and experimental use.

The process developed here included assessment of scores based upon a panel of 9 people, which provided standard deviation values for each of the scores. This range of values provided mean scores from the team, and the deviation values were consistent with others who have conducted FMEA[22]. The origins of these can be a result of the variability in the experiences of the physicists, biomedical engineers, researchers, and PhD student, but generally are thought to be important to assess. While multiple perspectives allowed identification of many of the potential failure modes, it also resulted in larger uncertainties in the scores. During the scoring meetings, there were different interpretations of the scoring guideline and these required further discussion to better align the scoring between team members. This process of educational debate was thought to be a positive side product of the FMEA process.

In response to this analysis, to mitigate some of the errors involving beam monitoring and control, the institution will implement a checklist[31] after conversion of the machine to ensure the beam gating mechanism is turned on and overdosing of samples/animals is avoided. The team always emphasized the importance of secondary dose checks, which for most of the FLASH experiments have involved confirmation of delivery with dose rate independent Gafchromic film. Radiation damage of the detectors was also a concern due to changes in measured dose by the dose controlling system, and so currently there is ongoing exploration of potential alternatives for beam monitoring and control. One potential unconventional method would be to measure the air scintillation, as shown in **Figure 6a**, to monitor beam output because the radiation damage of the detector could be minimal and it would not be directly irradiated by the beam, while any detector in the beam is more susceptible to damage. Other potential detectors that may be viable are ionization chambers with shifted voltage potential and/or correction for ion recombination. Systems such as this are under development and could be used if they exhibit minimal radiation damage.

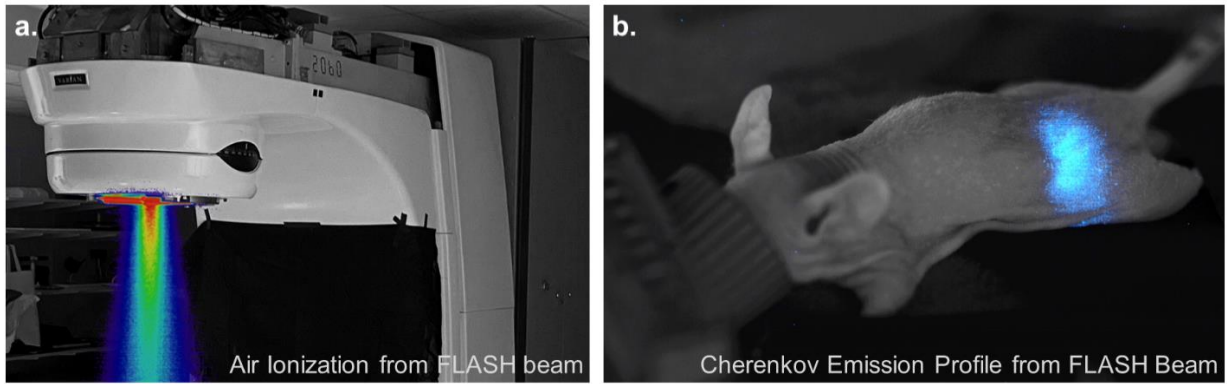

**Figure 6. a.** Air scintillation produced by the UHDR beam from a converted LINAC via imaging with an intensified CMOS camera. **b.** Cherenkov emission surface profile from FLASH irradiation of a mouse with a 2 cm diameter cutout imaged with an intensified CMOS camera

From discussion with the clinical physics team, there was a push to incorporate surface guidance during FLASH beam delivery as another verification of delivered treatments. Cherenkov emission imaging of patients is currently integrated into the clinic at the institution[32], and it has been demonstrated that it can be imaged at the single pulse resolution[30], so imaging in vivo Cherenkov under FLASH conditions is being explored for minimizing errors involving patient/sample set up[33] as shown in **Figure 6b**.

The biomedical engineers also suggested that sensors could be installed to ensure the positioning of the carousel, target, and energy switch after converting the machine to and from FLASH and would be ideal to minimize errors during conversion of the LINAC. While this has not been developed to date, certainly this could be done and likely should be done in commercial solutions.

While there were five failure modes with RPN>100, there were several other failure modes with RPN close to 100, when accounting for the standard deviation in the responses from the team. For example, the failure modes associated with turning of the servos circuit in **Table 2** were within two standard deviations of 100. Another source of failure modes would be the verification of consistent beam parameters (including data acquisition parameters and ambient noise in **Table 2**). To mitigate errors associated with the servos circuit and beam consistency measures, the checklist suggested by the physicists can be implemented. For beam measurements, the team also suggested automating data acquisition methods with the detector with minimal input from the user (e.g. automated data acquisition parameters, and online method of ambient noise reduction).

A final thought is that while other institutions may have different protocols or different FLASH irradiators, it may be advantageous to include an open-source survey for other physicists and professionals involved in FLASH studies. Such an open survey could include the failure modes identified in this study and give the opportunity to score RPN values specific to other institutions or experiences. This may better inform the wider risks and mitigations needs in the community, which is important given that FMEA is still a subjective assessment process, which has potential biases in scoring and inherent variability. With a larger sample of researchers, engineers, and physicists across the radiation oncology community, the scoring variability may be reduced with greater accuracy of the mean. If the open-source survey provided opportunity to include other potential failure not defined by this study, there might be other error mitigations that can improve and ensure safe delivery of FLASH beams. The survey may also inform researchers from other institutions about the conversion process, use of a LINAC for FLASH experiments, and the FMEA process. The exposure to FMEA via an online scoring survey can also promote its integration for other treatment modalities.

## 5. Conclusions

In this study, the potential errors of utilizing a clinical LINAC to deliver UHDR beams for FLASH experiments have been identified, with the experience of converting the machine nearly 100 times over the time course of more than a year. The team of 9 professional identified a total of 46 potential failure modes including 5 with RPN>100. These failure modes involved 1) patient set up, 2) gating mechanism in the beam delivery, and 3) the detector of the beam stopping mechanism. From the FMEA, some future methods to mitigate errors included 1) implementation of a checklist post conversion, 2) investigation of robust radiation detectors with reduced long-term damage from UHDR beams, 3) automation of the quality assurance and beam consistency checks, and 4) implementation of surface guidance during treatment delivery. For greater accuracy in the scoring and reduced variability, the FMEA of type of operation could potentially be opened to the larger radiation oncology community via an online survey, where they can score and provide other potential errors.

**Acknowledgments:** This work was supported by the Norris Cotton Cancer Center seed funding through core grant P30 CA023108 and through the shared irradiation service, as well as through seed funding from the Thayer School of Engineering, and from grant R01 EB024498. Department of Medicine Scholarship Enhancement in Academic Medicine (SEAM) Awards Program from the Dartmouth Hitchcock Medical Center and Geisel School of Medicine also supported this work.

**Disclosure of Conflicts of Interest:** Dr. Brian W. Pogue is a cofounder and Mahbubur Rahman is an employee of DoseOptics LLC, outside of the submitted work.